\documentclass[aps,prl,reprint,groupedaddress,longbibliography,superscriptaddress]{revtex4-2}
\usepackage{placeins}
\usepackage{amsmath,amsfonts,amssymb}
\usepackage{amsmath}
\usepackage{graphicx}
\usepackage[colorlinks=true, allcolors=blue]{hyperref}
\usepackage{siunitx}
\usepackage[defaultcolor=red,authormarkup=none,final]{changes}
\usepackage{soul}
\usepackage{comment}
\usepackage{layouts}
\graphicspath{{Figures/}}

\definechangesauthor[name={Dmitry Bykov}, color=red]{DB}
\definechangesauthor[name={Fewer Words}, color=purple]{FW}

\begin{document}
	\title{Ultra-high quality factor of a levitated nanomechanical oscillator}
	\author{Lorenzo Dania}
	\email[]{ldania@ethz.ch}
	\altaffiliation[Present address: ]{Photonics Laboratory, ETH Zürich, CH 8093 Zürich, Switzerland.}
	\affiliation{Institut f{\"u}r Experimentalphysik, Universit{\"a}t Innsbruck, Technikerstra\ss e 25, 6020 Innsbruck,
		Austria}
	\author{Dmitry S. Bykov}
	\email[]{dmitry.bykov@uibk.ac.at}
	\affiliation{Institut f{\"u}r Experimentalphysik, Universit{\"a}t Innsbruck, Technikerstra\ss e 25, 6020 Innsbruck,
		Austria}
	\author{Florian Goschin}
	\affiliation{Institut f{\"u}r Experimentalphysik, Universit{\"a}t Innsbruck, Technikerstra\ss e 25, 6020 Innsbruck,
		Austria}
	\author{Markus Teller}
	\altaffiliation[Present address: ]{ICFO-Institut de Ciencies Fotoniques, The Barcelona Institute of Science and Technology, 08860 Castelldefels (Barcelona), Spain}
	\affiliation{Institut f{\"u}r Experimentalphysik, Universit{\"a}t Innsbruck, Technikerstra\ss e 25, 6020 Innsbruck,
		Austria}
	\author{Abderrahmane Kassid}
	\affiliation{Physics Department, Ecole Normale Sup\'{e}rieure, 24 rue Lhomond, 75005 Paris, France}
	\author{Tracy E. Northup}
	\affiliation{Institut f{\"u}r Experimentalphysik, Universit{\"a}t Innsbruck, Technikerstra\ss e 25, 6020 Innsbruck,
		Austria}
	\date{\today}

\begin{abstract}
A levitated nanomechanical oscillator under ultra-high vacuum (UHV) is highly isolated from its environment. It has been predicted that this isolation leads to very low mechanical dissipation rates. However, a gap persists between predictions and experimental data. Here, we levitate a silica nanoparticle in a linear Paul trap at room temperature, at pressures as low as \SI{7e-11}{\milli\bar}. We measure a dissipation rate of 
\added{$2\pi\times\SI{69(22)}{\nano\hertz}$}, 
corresponding to a quality factor exceeding $10^{10}$, more than two orders of magnitude higher than previously shown. A study of the pressure dependence of the particle's damping and heating rates provides insight into the relevant dissipation mechanisms. 
\end{abstract}

\maketitle

The center-of-mass motion of a silica nanoparticle has recently been cooled to the quantum ground state~\cite{delic2020cooling,magrini2021realtime,tebbenjohanns2021quantum,kamba2022optical,ranfagni2022twodimensional,piotrowski2023simultaneous}, opening up the possibility to prepare nonclassical motional states of levitated objects consisting of billions of atoms~\cite{gonzalezballestero2021levitodynamics}. However, a prerequisite to prepare and analyze such exotic states---and to exploit them for applications in sensing, transduction, or tests of fundamental physics---is a low-dissipation environment that preserves quantum coherence.
For clamped nanomechanical oscillators, dissipation has been suppressed through engineering geometry and strain, 
resulting in quality factors above $10^{10}$~\cite{tsaturyan2017ultracoherent,maccabe2020nanoacoustic,beccari2022strained,bereyhi2022perimeter}.
For levitated objects, in contrast, collisions with background-gas molecules are typically the dominant source of dissipation~\cite{epstein1924resistance,beresnev1990motion,cavalleri2010gas}, and thus the route to suppressing dissipation lies not in materials engineering but in reducing the pressure.
It has been estimated that damping rates of $2\pi\times\SI{200}{\nano\hertz}$ and quality factors of $\SI{3e12}{}$ can be achieved in UHV~\cite{chang2010cavity}.
		
Direct measurements of dissipation are crucial for levitated particles because additional heating and damping mechanisms become important as the pressure is reduced; it can no longer be assumed that gas damping dominates. Light scattering is of particular concern: for a particle confined in an optical tweezer, dissipation due to radiation damping already dominates over gas damping at high-vacuum pressures~\cite{ashkin1976optical,hinkle1990pressure,li2011millikelvin,jain2016direct}.
Quality factors of $\SI[print-unity-mantissa=false]{e8}{}$ have been reported for an optically trapped nanoparticle~\cite{gieseler2013thermal}.
Electrical and magnetic traps are not compromised by light-induced decoherence, and for nanoparticles in such traps, quality factors of \SI{2.6e7}{} and dissipation rates of $2\pi\times\SI{0.59}{\micro\hertz}$ have been reported~\cite{slezak2018cooling,vinante2020ultralow,pontin2020ultranarrowlinewidth,leng2021mechanical,hofer2022highq}.

Here, using a nanoparticle confined in a linear Paul trap in UHV, we measure an ultra-low dissipation rate of \added{$2\pi\times\SI{69(22)}{\nano\hertz}$} and a record quality factor of \added{\SI{1.8(6)e10}{}}. These values pave the way for detection of forces~and electric fields with sensitivities comparable to or surpassing the current benchmarks set by trapped atoms~\cite{millen2020optomechanics,moore2021searching}. Moreover, the low pressure demonstrated here is a key condition to test wavefunction-collapse models~\cite{goldwater2016testing,vinante2019testing} and to engineer long-lived quantum states of motion of macroscopic mechanical oscillators~\cite{chang2010cavity,Romero-Isart2010,romeroisart2011large}. 
In light of these goals, we characterize the noise environment in UHV, identifying the most likely noise sources.

	\begin{figure}[ht]
	\centering
	\includegraphics[width=1\linewidth]{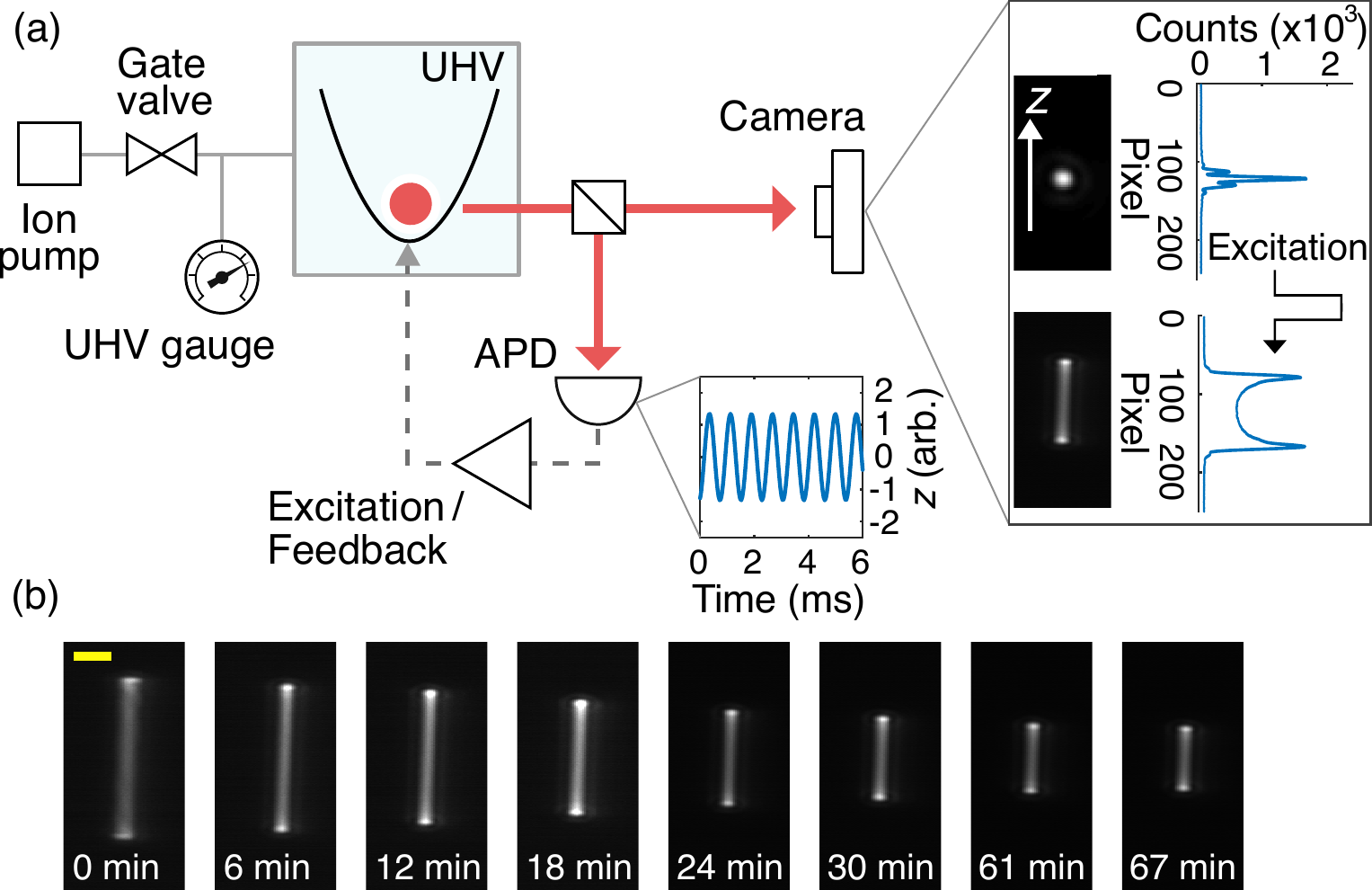}
	\caption{Particle trapping and detection. (a) Schematic of the ultra-high-vacuum (UHV) setup and detection schemes. A nanoparticle is detected in two ways: it is imaged with a camera, and its position is measured with an avalanche photodiode (APD) using a confocal technique. The camera inset shows particle images before and after excitation, as well as position histograms. The APD inset shows a time trace of the particle's position. (b) Camera snapshots of the particle with decaying oscillation amplitude at $P=\SI{5.4e-8}{\milli\bar}$. Scale bar: $\SI{20}{\micro\meter}$.}
	\label{fig:fig_1}
\end{figure}

Our primary goal is to investigate damping of the particle's motion as a function of pressure.
\added[id=FW]{Experiments are conducted} in a vacuum chamber in which we vary the pressure from \SI{7e-11}{\milli\bar} to  \SI[print-unity-mantissa=false]{1e-4}{\milli\bar}.
To increase the pressure from its minimum value, we limit the effective pumping speed of the chamber's combination pump (a non-evaporable getter and an ion pump) by partially closing a gate valve.
With the valve fully closed, we reach \SI[print-unity-mantissa=false]{1e-4}{\milli\bar} in one day.
To decrease the pressure back to UHV, we first \added[id=FW]{reach \SI[print-unity-mantissa=false]{e-9}{\milli\bar}}
with a turbomolecular pump. We then open the gate valve and pump with the combination pump to below $\SI[print-unity-mantissa=false]{e-10}{\milli\bar}$. \added{With this room-temperature setup, we expect that pressures one order of magnitude lower can be reached~\cite{obsil2019roomtemperature}.}

The silica particle used for the measurements presented here was loaded at \SI{e-9}{\milli\bar} via laser-induced acoustic desorption and temporal control of the Paul-trap potential~\cite{bykov2019direct}; these methods allow us to load particles over a range of pressures down to UHV. 
The particle has charge $q = +\SI{300(30)}{\elementarycharge}$ and a nominal diameter of \SI{300}{\nano\meter}; however, since we measure a mass $m=\SI{4.3(4)e-17}{\kilo\gram}$ that is twice the mass of particles previously loaded from the same source~\cite{dania2021optical,dania2022position}, it is likely that the particle is a cluster of two nanospheres with a somewhat larger size.
The particle's motion is detected optically (Fig.~\ref{fig:fig_1}a): 
a \SI{780}{\nano\meter} laser beam with a power of \SI{21}{\milli\watt}
is focused with a waist of \SI{300}{\micro\meter} on the particle; a lens collects the scattered light.  
Two methods are used to extract information from the collected light about the particle's motion:
confocal detection realized with a fiber-coupled avalanche photodiode (APD)\cite{vamivakas2007phasesensitive,kuhn2015cavityassisted,xiong2021lensfree,bykov2023sympathetic}, and imaging of the particle on a CMOS camera~\cite{leng2021mechanical}.
We use the APD method to track the particle's position in real time.
A typical APD signal is shown in an inset of Fig.~\ref{fig:fig_1}a.
With the camera, we measure the oscillation amplitude averaged over the camera acquisition time. Compared to the APD method, the camera method has lower spatial and temporal resolution but offers a larger field of view,
allowing us to measure particle amplitudes up to hundreds of micrometers.  
Moreover, the camera detection is more resilient to drifts in the optics alignment and laser power, making it possible 
to perform days-long measurements.

To modify the particle's motion, we apply electrical forces by supplying a suitable voltage to the trap electrodes~\cite{dania2021optical}.
To reduce the particle's amplitude, we apply feedback based on the real-time APD detection; to increase the amplitude, we apply a sinusoidal driving force at the mechanical resonance frequency of the trapped particle. An inset in Fig.~\ref{fig:fig_1}a shows images of the feedback-cooled particle and the driven particle.

For measurements of the particle's damping, we focus on motion along the trap's $z$ axis, for which DC voltages provide confinement. We choose this axis because here the particle is less susceptible to noise of the AC trap drive~\cite{roos1999quantum}. The equation of motion
	$\ddot{z}+\gamma\dot{z}+\Omega_z^2z=\frac{1}{m}\mathcal{F}_{\text{th}}$~\cite{gieseler2014dynamic} 
describes the particle's position, where $\gamma$ is the damping rate, which is proportional to the background pressure $P$~\cite{beresnev1990motion,jain2016direct}; $\Omega_z/2\pi=\SI{1.28}{\kilo\hertz}$ is the oscillation frequency; and $\mathcal{F}_{\text{th}}$ is the stochastic force due to thermalization with the environment.
Here we assume that $\mathcal{F}_{\text{th}}$ dominates over all other stochastic forces, such as those due to laser, electronic, and displacement noise.

We determine $\gamma$ using two methods.
The first method, ring-down, consists of measuring the amplitude relaxation
\begin{equation}
	\langle z(t)^2\rangle=\langle z(0)^2\rangle e^{-\gamma t}
	\label{eq:ringdown},
\end{equation}
where Eq.~\ref{eq:ringdown} holds when the viscous force dominates over $\mathcal{F}_{\text{th}}$, such that the particle follows a deterministic trajectory. 
Figure~\ref{fig:fig_1}b shows \added[id=FW]{images from} a relaxation experiment. \added{While a particle with large motional amplitude is prone to trap anharmonicities, the ring-down method is insensitive to them~\cite{stipe2001noncontact}.} The second method, ring-up, consists of preparing the particle at a low temperature $T_\text{fb}$ via feedback cooling, then switching off cooling and measuring the particle's thermalization with background gas at room temperature $T_\text{0}$.
When $\mathcal{F}_{\text{th}}$ dominates over the viscous force, the particle follows a stochastic trajectory in time. Averaging over an ensemble of such trajectories allows us to obtain the \added[id=FW]{mean} energy~\cite{gieseler2014dynamic}
\begin{equation}
	\langle E(t)\rangle=k_{\text{B}}T_{\text{0}}+k_{\text{B}}(T_{\text{fb}}-T_{\text{0}})e^{-\gamma t},
	\label{eq:reheating}
\end{equation}
from which we extract $\gamma$.
We use the ring-down method because it allows us to characterize the damped harmonic oscillator in a single measurement. We use the ring-up method for two reasons: first, a fit to Eq.~\ref{eq:reheating} for data taken under low vacuum (i.e., in thermal equilibrium) establishes a calibration between the APD voltage and the energy $k_{\text{B}}T_{\text{0}}$. \added{We use this calibration to obtain $T_\text{fb}$ ~\cite{Note1}.} Second, we compare the heating rate determined from the ring-up method with the rate expected from thermal noise, in order to determine whether non-thermal noise sources play a significant role at UHV.
\begin{figure}[!ht]
	\includegraphics[width=1\linewidth]{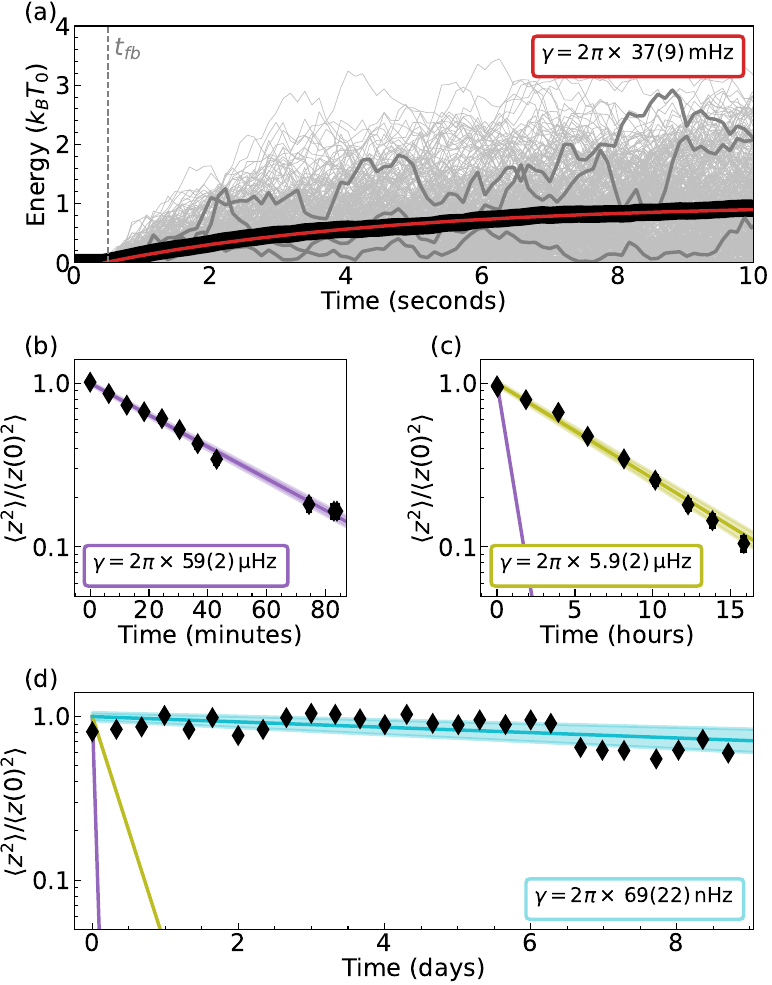}

	\caption{Measurements of the damping rate $\gamma$. (a) Ring-up measurement at $P_1=\SI{1.2e-4}{\milli\bar}$.  Gray: \added[id=FW]{400 individual} data traces. \added[id=FW]{Three examples are highlighted in dark gray.} Black: ensemble average. Red: fit of the ensemble-averaged data with Eq.~\ref{eq:reheating}, where $T_{\text{fb}}$ is fixed to \SI{1}{\kelvin}. The dashed line $t=t_{\text{fb}}=\SI{0.5}{\second}$ indicates the time at which feedback cooling is switched off. (b) Ring-down measurement at $P_2=\SI{5.4e-8}{\milli\bar}$, (c) $P_3=\SI{5e-9}{\milli\bar}$, and (d) $P_4=\SI{7e-11}{\milli\bar}$, with a logarithmic scale used for the ordinate axis. Error bars on the data are smaller than the diamond symbols; they correspond to the uncertainty in determining amplitudes from camera images. Solid lines represent fits with Eq.~\ref{eq:ringdown} \added{in logarithmic scale, with $\langle z(0)^2\rangle$ and $\gamma$ as fit parameters.} \added[id=FW]{Light shaded regions indicate} $1\sigma$ error bars on fit parameters. Fits for higher pressures are reproduced in the lower-pressure plots for comparison.}
	\label{fig:fig_2}
\end{figure}

\begin{figure}[ht]
	\includegraphics[width=1\linewidth]{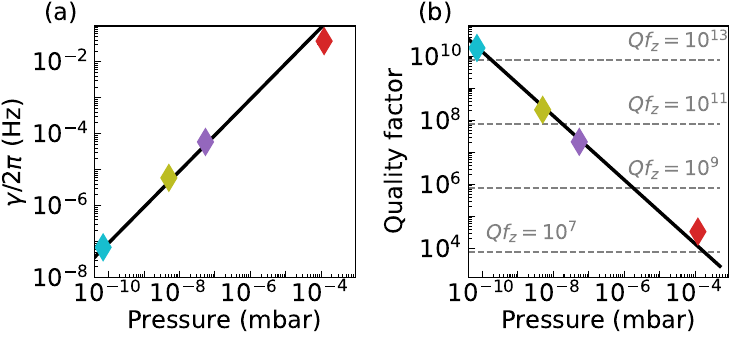}
	\caption{Quality factor of the levitated oscillator. (a) Damping rate $\gamma$ as a function of pressure. Error bars for $\gamma$ represent the uncertainty of the fit parameter, while error bars for pressure represent the imprecision of the pressure gauge. Error bars are smaller than the diamond symbols. Colors indicate the corresponding fits in Fig.~\ref{fig:fig_2}. The solid line shows a linear fit $\gamma/(2\pi) = a P$ to the data in logarithmic scale, where the parameter $a$ depends on the particle and the background gas properties~\cite{beresnev1990motion,jain2016direct}. \added{Since non-negligible uncertainties are present in both $\gamma$ and $P$, we use a total least squares regression to fit the data~\cite{golub1980analysis}.}
	The fit yields \added{$a=\SI{0.9(2)}{\kilo\hertz\per\milli\bar}$. See Ref.~\cite{Note1} for a comparison with theory.}
	(b) Quality factor at four pressures. Error bars are propagated from the uncertainties of $\gamma$ shown in (a). The solid line represents the function $Q = \Omega_z / (2\pi aP)$, where $a$ is obtained from the fit in (a). Dashed lines indicate values for the product $Qf_z$, where $f_z=\Omega_z/(2\pi)$ is the oscillation frequency.}
	\label{fig: linewidth vs. pressure}
\end{figure}
In a first set of measurements, we analyze how $\gamma$ changes when $P$ is varied. 
We extract $\gamma$ with a ring-up measurement at $P_1 = \SI{1.2e-4}{\milli\bar}$ and with ring-down measurements at $P_2 = \SI{5.4e-8}{\milli\bar}$, $P_3=\SI{5e-9}{\milli\bar}$, and $P_4=\SI{7e-11}{\milli\bar}$.
At $P_1$, the particle is prepared at $T_{\text{fb}}=\SI{1}{\kelvin}$ and monitored with the APD \added[id=FW]{during thermalization} for $\SI{10}{\second}$. 
This sequence is repeated 400 times.
For each repetition, we square the position data and average \added[id=FW]{them} over \SI{0.1}{\second}. This procedure determines the particle's energy with a \added[id=FW]{\SI{10}{\hertz}} sampling rate, much faster than the expected ring-up rate of $\SI[print-unity-mantissa=false]{\sim e-2}{\hertz}$.
In Fig.~\ref{fig:fig_2}a, we plot the energy traces together with their ensemble average. We then fit the ensemble average with Eq.~\ref{eq:reheating}, under the assumption that the particle thermalizes at $T_0 = \SI{300}{\kelvin}$~\cite{bykov2019direct, gieseler2014dynamic,pontin2020ultranarrowlinewidth}. From this fit, we extract  \added{$\gamma_{P_1}=2\pi\times\SI{37(9)}{\milli\hertz}$}. 

For the ring-down measurements at pressures $P_2$, $P_3$, and $P_4$, the particle's motion is initially excited to an amplitude of hundreds of microns, and the subsequent relaxation is measured with gated camera imaging. The particle is illuminated only for the camera measurement time, $\SI{1}{\second}$ for each acquisition, so that the particle's motion is minimally influenced by the laser, e.g., due to radiation damping or force noise induced by power fluctuations~\cite{jain2016direct}. 
Figures~\ref{fig:fig_2}b, \ref{fig:fig_2}c, and \ref{fig:fig_2}d show the normalized squared amplitude of the particle's motion as a function of time, measured at $P_2$, $P_3$, and $P_4$. 
A fit of the data with Eq.~{\ref{eq:ringdown}} yields \added{$\gamma_{P_2}=2\pi\times\SI{59(2)}{\micro\hertz}$,
$\gamma_{P_3}=2\pi\times\SI{5.9(2)}{\micro\hertz}$} and 
\added{$\gamma_{P_4}=2\pi\times\SI{69(22)}{\nano\hertz}$}.
At the lowest pressure, $P_4$, the uncertainties associated with camera detection are much smaller than the residuals of the exponential fit, indicating that Eq.~\ref{eq:ringdown} does not fully capture the particle's dynamics~\cite{Note1}.
This would occur if the viscous force were not dominant. The viscous force is three orders of magnitude larger than $\mathcal{F}_{\text{th}}$ under the assumption that $\mathcal{F}_{\text{th}}$ is due to background-gas collisions at room temperature.
However, we cannot rule out the possibility that $\mathcal{F}_{\text{th}}$ is dominant: under high vacuum, the equilibrium temperature of a nanoparticle in a Paul trap has been shown to vary inversely with pressure due to noise on the trap electrodes~\cite{pontin2020ultranarrowlinewidth}; furthermore, the noise seen by a particle oscillating over hundreds of micrometers may be significantly higher than for a localized particle~\cite{penny2022lownoise}. A second possibility is that the particle's dynamics were disrupted by infrequent spikes in electronic or vibrational noise, which could explain the bimodal distribution of the residuals.

In Fig.~\ref{fig: linewidth vs. pressure}a, we plot \added[id=FW]{$\gamma$} determined from the fits of Fig.~\ref{fig:fig_2} \added[id=FW]{for different pressures}.
The data are consistent with a linear model $\gamma \propto P$, from which we infer that at pressures as low as \added[id=FW]{$P_4$}, damping is still dominated by \added[id=FW]{background-gas} collisions.
In Fig.~\ref{fig: linewidth vs. pressure}b, we plot the quality factors $Q=\Omega_z/\gamma$. The highest value for \added[id=FW]{$Q$}, obtained at $P_4$, is \added{$1.8(6)\times10^{10}$}, \added[id=FW]{corresponding} to \added{$2.4(7)\times10^{13}$} for the $Q$-frequency product, a benchmark for optomechanics in the quantum regime~\cite{aspelmeyer2014cavity}. 
\added{A second set of data at $P_4$ yields $Q=\SI{3(2)e10}{}$ for different trap parameters~\cite{Note1}.}

\begin{figure}[ht]
	\includegraphics[width=1\linewidth]{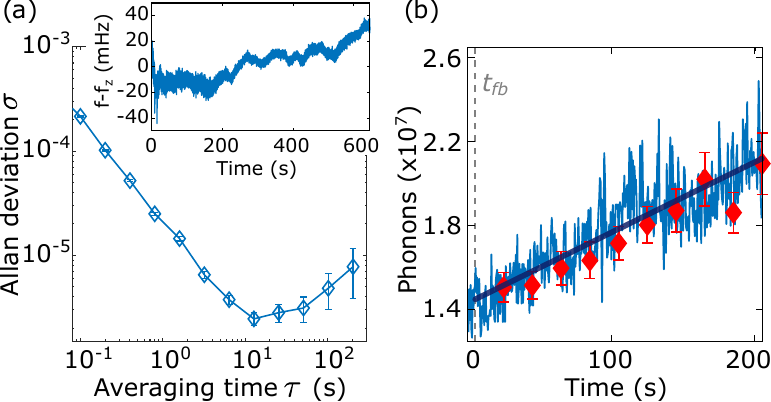}
	\caption{(a) Allan deviation of the particle oscillation frequency at pressure $P_4$. Lines connect neighboring data points. Error bars represent the standard error of the frequency mean. Inset: frequency drift as a function of time. (b) Heating-rate measurements at $P_4$ for continuous (blue) and stroboscopic (red) illumination of the particle with the detection laser. Error bars on the stroboscopic measurements represent the standard deviation of individual temperature measurements. The dashed line $t=t_{\text{fb}}=\SI{5}{\second}$ indicates the time at which feedback cooling is switched off. The dark-blue line is a fit to the data for continuous illumination, the slope of which yields the heating rate $\Gamma_{\rm tot}=\SI{3.3(2)e4}{\text{phonon}/\second}$. \added{The fit for stroboscopic illumination is not shown since it coincides with the fit for the continuous case.}}
	\label{fig:allan plot}
\end{figure}

The ultra-high quality factors and low \added[id=FW]{dissipation}  demonstrated here open up the possibility to use levitated particles in Paul traps for force sensing and tests of quantum mechanics.
For such applications, it is important to quantify both the frequency stability and the heating rate of the nanoparticle oscillator in \added[id=FW]{UHV}.
\added[id=FW]{To this end,} we calculate the Allan deviation to characterize the frequency stability.
We record a \added{time trace of the particle's position for $t_{\mathrm{f}}=\SI{600}{\second}$} and extract the particle frequency as a function of time via a phase-locked loop in post-processing. For this measurement, the particle evolves freely in the Paul-trap potential at pressure~\added[id=FW]{$P_4$}.
From the discrete frequency time-trace, we determine the Allan deviation as 
\begin{equation}
\sigma(\tau)=
\left[\frac{1}{2f_z^2} \frac{1}{\added{N(\tau)}-1} \sum_{k=2}^{\added{N(\tau)}}(\bar{f}_{k} - \bar{f}_{k-1})^2\right]^{1/2},
\end{equation} 
with $f_z=\Omega_z/(2\pi)$\added{; $N(\tau) = \left[t_{\mathrm{f}}/\tau\right]$ the number of intervals, where $[a/b]$ represents integer division; } and $\bar{f}_k$ the average frequency in the $k\text{th}$ interval of duration~$\tau$. The results are shown in Fig.~\ref{fig:allan plot}a. For an optimum averaging time \added{$\tau_\textrm{opt}$}$=\SI{20}{\second}$, a fractional frequency fluctuation of $\sigma(\added{\tau_\textrm{opt}}) =\SI{2e-6}{}$ is achieved.
\added{For comparison, one expects a thermally limited value~\cite{gavartin2013stabilization} of $\sigma(\tau_\textrm{opt})=1/\sqrt{Q\Omega_z\tau_\textrm{opt}}=\SI{2e-8}{}$, which we do not reach due to inefficient detection~\cite{manzaneque2023resolution} and mechanical frequency fluctuations.}
For larger values of $\tau$, $\sigma$ increases due to drifts of the particle's frequency. We extract a linear drift rate of $\SI{\sim8e-8}{\hertz\per\second}$.
Possible causes of the drift include nonlinearities of the confining potential, fluctuations in trap-electrode voltages, and fluctuations in the detection-laser power.

In a final set of measurements, we determine the \added[id=FW]{nanoparticle's} heating rate to quantify the noise environment\added{, which is assumed to be white over the ultra-narrow bandwidth of the oscillator.}
We carry out a ring-up measurement at \added[id=FW]{$P_4$}: 
The particle is prepared at $T_{\rm fb}=\SI{0.8}{\kelvin}$\added{, which we assume corresponds to a thermal state.} Next, feedback cooling is turned off and the particle's position is monitored for $\SI{200}{\second}$ with continuous laser illumination. This sequence is repeated $\SI{100}{}$ times.
As the $1/e$ ring-up time expected from $\gamma_{P4}$ is approximately one month, we expand Eq.~\ref{eq:reheating} to first order in time 
\begin{align}
	\langle E(t)\rangle &= k_{\text{B}} T_{\text{fb}} + k_{\text{B}}(T_{\text{0}}- T_{\text{fb}}) \gamma t \nonumber \\
	&\approx k_{\text{B}} T_{\text{fb}} + k_{\text{B}}T_{\text{0}} \gamma t, 
	\label{eq: linear}
\end{align}
which corresponds to a heating rate $\Gamma_{\text{gas}}= k_BT_0 \gamma/(\hbar\Omega_z)$, \added[id=FW]{assuming} that the force noise is dominated by gas collisions.
From the averaged data (Fig.~\ref{fig:allan plot}b), we extract a rate \added{$\Gamma^{\rm bright}_{\rm meas}$}$=\SI{3.3(2)e4}{\text{phonon}/\second}$.
In comparison, our ring-down measurement predicts a rate \added{$\Gamma_{\text{gas}}= k_BT_0 \gamma_{P4}/(\hbar\Omega_z)=\SI{2.1e3}{\text{phonon}/\second}$, $16$} times smaller than our measured value. 

From the ring-up measurement, we infer that background-gas recoil is not the dominant noise source.
One possible noise source is the detection laser; to exclude this,
we re-run the heating-rate experiment in the dark.
In this case, the laser is turned on only during the first $\SI{5}{\second}$ of each ring-up in order to initialize the particle via feedback cooling, and for $\SI{500}{\milli\second}$ every $\SI{20}{\second}$ in order to stroboscopically measure the particle's energy during its free evolution.
The results of this measurement are also plotted in Fig.~\ref{fig:allan plot}b; we extract a rate \added{$\Gamma^{\rm dark}_{\rm meas}$}$=\SI{3.1(8)e4}{\text{phonon}/\second}$,  consistent with the rate determined under continuous illumination, from which we deduce that the laser does not play a significant role.

A second possible stochastic-force source is electric-field noise, due to fluctuations in AC and DC drive fields, Johnson noise, or coupling between the particle's secular motion and micromotion. 
From \added{$\Gamma^{\rm dark}_{\rm meas}$}, we extract a white force-noise spectrum $S_{ff}=4m\hbar\Omega_z\added{\Gamma^{\rm dark}_{\rm meas}}=\SI{4e-42}{\newton^2\hertz^{-1}}$ acting on the particle. If $S_{ff}$ is dominated by electric-field noise, we can determine the electric-field-noise spectrum:
$S_{EE}=S_{ff}/q^2=\SI{1.7e-9}{(\volt\per\meter)^2\hertz^{-1}}$, equivalent to an electronics voltage noise of $S_{v}=$ \added{$d\sqrt{S_{EE}}=$} $ \SI{38}{\nano\volt\per\sqrt{\hertz}}$\added{, where $d=\SI{0.92}{\milli\meter}$ is the particle-to-electrode distance}. 
\added[id=FW]{For} comparison, Vinante et al. estimate that $S_v = \SI{10}{\nano\volt\per\sqrt\hertz}$ is around the threshold of what can be achieved without extraordinary engineering efforts~\cite{vinante2019testing}. Thus, if the electric field is the dominant noise source, we have room for improvement. \added[id=FW]{But} reducing $S_v$ to this threshold value would still yield a heating rate slightly larger than $\Gamma_{\rm gas}$; to suppress electric-field noise further---in particular, for future experiments at even lower pressures---we could reduce $q$.

A third possibility is displacement noise due to trap vibrations. If $S_{ff}$ is dominated by displacement noise~\cite{weiss2021large}, we infer a vibration energy density $S_{zz}=2\hbar\added{\Gamma^{\rm dark}_{\rm meas}}/(\pi m\Omega_z^3)=\SI{9.5e-26}{\meter^2\hertz^{-1}}$. \added{We have measured $S_{zz}$ at the vacuum chamber to be $\SI{2e-25}{\meter^2\hertz^{-1}}$, suggesting that displacement noise may be dominant~\cite{Note1}.}
To discriminate between electric and displacement noise, it \added[id=FW]{will} be sufficient to study the dependence of \added{$\Gamma^{\rm dark}_{\rm meas}$} on $q$, since \added{$\Gamma^{\rm dark}_{\rm meas}$} is independent of $q$ for displacement noise. It has been argued that vibrational noise can be suppressed to negligible values in experiments with levitated nanoparticles~\cite{vinante2019testing}. \added{Two other possible noise sources are discussed in Ref.~\footnote{See Supplemental Material at [URL will be inserted by publisher] for the ion-trap geometry, detector calibration, a comparison of ring-down and ring-up measurements, analysis of residuals, discussions of noise due to electrode resistivity and coupling to rotational or librational modes, vibration measurements, analysis of the particle frequency, an estimate of the particle damping coefficient from theory, and additional ring-down data. The Supplemental Material includes Refs.~\cite{Blatt_RFheating,martinetz2022surfaceinduced,lide1995crc,kumph2016electricfield,DAHNEKE,ganta2011optical,millen2014nanoscale}.}.}

In conclusion, with a levitated nanoparticle, we have realized a mechanical oscillator with a quality factor of \added{\SI{1.8(6)e10}{}}, determined from the damping rate of the particle's oscillations at \SI{7e-11}{\milli\bar}. At this pressure, the \added[id=FW]{calculated} collision rate with background-gas molecules is $\SI{1.1}{\kilo\hertz}$, that is, on average, one molecule collides with the particle every 1.2 oscillation cycles. For levitated optomechanical experiments in the quantum regime, it will be necessary to operate at similar or lower pressures~\cite{romeroisart2011large,bose2017entanglement,vinante2019testing,weiss2021large,neumeier2022quantum}, which is facilitated by the direct loading method demonstrated here. 
While the measured damping rates are consistent with a gas-damping model~\cite{beresnev1990motion,jain2016direct}, the measured heating rates are higher than expected solely from interaction with the background gas; we have identified the most likely noise sources and routes to address them.

One application of this work is ultrasensitive force detection~\cite{gonzalezballestero2021levitodynamics,millen2020optomechanics}:
Force-noise and damping-rate measurements in UHV will allow wavefunction-collapse models to be tested in unexplored regimes~\cite{zheng2020room,pontin2020ultranarrowlinewidth,carlesso2022present}. \added{Pulsed-measurement schemes such as the one adopted here will be necessary to exploit these high quality factors.} Furthermore, if we introduce self-homodyne detection~\cite{dania2022position} and cold damping, we expect to prepare the particle's motional ground state within the dark potential of the electrodynamic trap, enabling tests of quantum physics in the absence of photon recoil.

Source data are available on Zenodo (\url{https://doi.org/10.5281/zenodo.10705895}).

\begin{acknowledgments}
We thank John Bollinger, Milena Guevara-Bertsch, and Marc Rodà Llordé for valuable discussions and Simon Baier for helpful feedback on the manuscript.
We acknowledge support for the research of this work from the Austrian Science Fund (FWF) grants I5540, W1259 and Y951.
\end{acknowledgments}
	
\bibliography{bibliography}
	
\end{document}


\title{Supplemental material for ``Ultra-high quality factor of a levitated nanomechanical oscillator"}
	\author{Lorenzo Dania}
	\email[]{ldania@ethz.ch}
	\altaffiliation[Present address: ]{Photonics Laboratory, ETH Zürich, CH 8093 Zürich, Switzerland.}
	\affiliation{Institut f{\"u}r Experimentalphysik, Universit{\"a}t Innsbruck, Technikerstra\ss e 25, 6020 Innsbruck,
		Austria}
	\author{Dmitry S. Bykov}
	\email[]{dmitry.bykov@uibk.ac.at}
	\affiliation{Institut f{\"u}r Experimentalphysik, Universit{\"a}t Innsbruck, Technikerstra\ss e 25, 6020 Innsbruck,
		Austria}
	\author{Florian Goschin}
	\affiliation{Institut f{\"u}r Experimentalphysik, Universit{\"a}t Innsbruck, Technikerstra\ss e 25, 6020 Innsbruck,
		Austria}
	\author{Markus Teller}
	\altaffiliation[Present address: ]{ICFO-Institut de Ciencies Fotoniques, The Barcelona Institute of Science and Technology, 08860 Castelldefels (Barcelona), Spain}
	\affiliation{Institut f{\"u}r Experimentalphysik, Universit{\"a}t Innsbruck, Technikerstra\ss e 25, 6020 Innsbruck,
		Austria}
	\author{Abderrahmane Kassid}
	\affiliation{Physics Department, Ecole Normale Sup\'{e}rieure, 24 rue Lhomond, 75005 Paris, France}
	\author{Tracy E. Northup}
	\affiliation{Institut f{\"u}r Experimentalphysik, Universit{\"a}t Innsbruck, Technikerstra\ss e 25, 6020 Innsbruck,
		Austria}
	\date{\today}
	\maketitle
	\onecolumngrid
	\appendix

\section{Ion-trap geometry}
\begin{figure}[ht]
	\centering
	\includegraphics[width=1\linewidth]{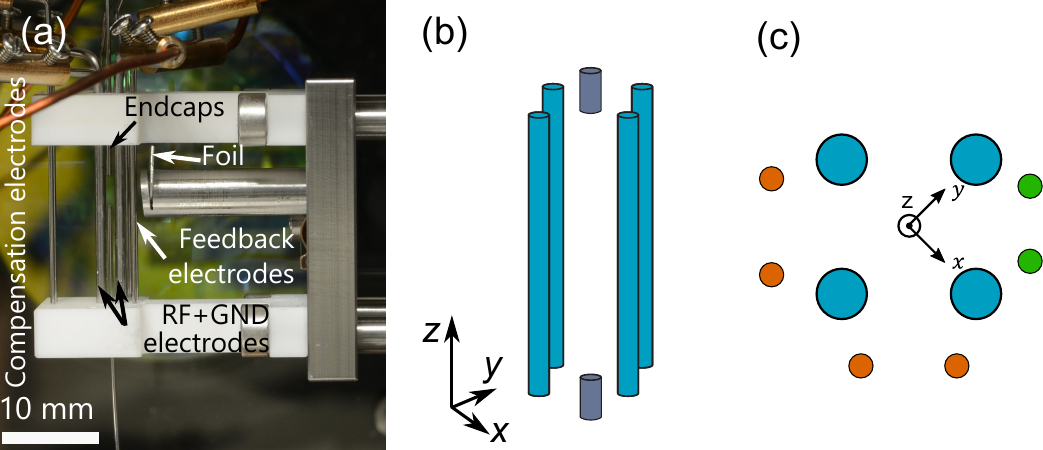}
	\caption{Paul trap geometry. (a) Photo of the Paul trap. The Paul-trap radiofrequency (RF), ground (GND), and endcap electrodes are indicated, along with electrodes for active feedback and micromotion compensation. The foil from which nanoparticles are loaded is also indicated. (b) Schematic side view of the Paul trap. Blue: RF electrodes. Grey: endcap electrodes. (c) Schematic top view of the Paul trap. Blue: RF electrodes. Orange: electrodes for micromotion compensation. Green:  electrodes for active feedback.}
	\label{figS1:trap_geometry}
\end{figure}

Fig.~\ref{figS1:trap_geometry}a shows a photo of the linear Paul trap, including electrodes for active feedback and micromotion compensation. A metal foil with nanoparticles is mounted in a cylindrical holder located on the right side of the trap. Fig.~\ref{figS1:trap_geometry}b shows a schematic side view of the Paul trap with the coordinate system used in the main text, namely, with the \textit{z}~axis along the trap axis. The compensation and feedback electrodes are not shown. Fig.~\ref{figS1:trap_geometry}c shows a schematic top view of the Paul trap, in this case with the compensation and feedback electrodes included but the endcap electrodes omitted.

\section{APD detector calibration via ring-up fitting}
\label{app:ring-up_calibration}
In this section, we explain how we calibrate data from the avalanche photodiode (APD) detector. 
This calibration is used to determine both the particle temperature under feedback cooling and the heating rates. 

The variance $\sigma$ of the signal from the APD is proportional to the particle's motional temperature. To convert $\sigma$ from units of \si{\volt^2} to units of \si{\joule}, we rely on the equipartition theorem. 
In the absence of feedback cooling, and under the assumption that the stochastic force due to thermalization dominates over all other stochastic forces, the particle motion thermalizes with the background gas at room temperature $T_0 = \SI{300}{\kelvin}$, as described by Eq.~2 in the main text. 
In practice, our calibration consists of a ring-up measurement at pressure $\sim\SI{1e-4}{\milli\bar}$, two examples of which are shown in Fig.~2a in the main text and in Fig.~\ref{fig:Guilio ringdown ringup}a here.
In both examples, feedback cooling is switched off at time $t = t_\mathrm{fb}$. 
To extract the calibration factor, we fit $\sigma(t)$ for $t>t_{\text{fb}}$ to the equation
\begin{equation}\label{eq:calibration_fit}
\sigma(t)=a+\left(b-a\right)e^{-\gamma t},
\end{equation} 
where $a$ and $\gamma$ are fit parameters and $b$ is defined as the average variance of the APD signal over the time for which feedback cooling is active. The model parameters are linked to temperature via $a=\alpha^2 k_BT_0$ and $b=\alpha^2 k_BT_{\text{fb}}$, where $\alpha$ is the proportionality coefficient between the APD voltage signal $u$ and the particle's displacement from the trap center $z$,
\begin{equation}
	u(t) = \alpha z(t),
	\label{eqn:proportionality}
\end{equation}
 and $T_{\text{fb}}$ is the particle temperature at the beginning of the ring-up. (Compare Eq.~\ref{eq:calibration_fit} with Eq.~2 in the main text.) Thus, to infer the particle energy in units of $k_BT_0$, where $k_B$ is the Boltzmann constant, we renormalize the APD variance with the fit parameter $a$: $\sigma\to\frac{1}{a}\sigma$. 
The temperature $T_{\text{fb}}$ at this pressure can be determined from $a, b$ and $T_0$: $T_{\text{fb}} = T_0 b/a$.
Note that, once in ultra-high vacuum (UHV), we cannot calibrate our APD detection since thermalization occurs on a time scale of weeks. We rely on the calibration done at $\sim\SI{1e-4}{\milli\bar}$ to quantify the temperature and heating rates in UHV.

This calibration is valid for all experimental data as long as the intensity of light collected from the particle does not vary. 
In order to apply the APD calibration at different laser powers, we rely on a second calibration technique. Prior to ring-up measurements, both in high vacuum and in UHV, we apply a periodic force to the particle by sending a voltage at frequency $\Omega_{\text{cal}}=\SI{300}{\hertz}$ to one of the endcap electrodes. The trapped particle responds by oscillating at $\Omega_{\text{cal}}$ with an amplitude $A=F/(m\Omega_z^2)$, where $m$ is the particle's mass; $\Omega_z$ is the particle's secular frequency in the Paul trap along the $z$ axis; and $F$ depends on the particle's charge, the distance of the endcap electrode from the particle, and the amplitude of the drive signal. Importantly, $A$ does not depend on the pressure. The amplitude $A$ is mapped to the APD signal amplitude via $A' = \alpha A$. We measure $A'$ by evaluating the APD signal in the frequency domain.
Each measured variance $\sigma$ is then rescaled by the factor $\left(A'_{\text{cal}}/A'_{\text{meas}}\right)^2$, where $A'_{\text{cal}}$ is the APD amplitude recorded before the calibration ring-up and $A'_{\text{meas}}$ is the APD amplitude recorded before the ring-up measurement at hand. 
For the data in Fig.~4b of the main text, the rescaling factor is $\left(A'_{\text{cal}}/A'_{\text{meas}}\right)^2 = 30.1$; a single measurement of $A'_{\text{meas}}$ was carried out before the ring-up sequence was repeated 100 times. For the data in Fig.~\ref{fig:Guilio ringdown ringup}a of Appendix~\ref{app:ring-up_ring-down}, $A'_{\text{meas}}$ was measured after every 40 ring-up sequences.

To determine $T_{\text{fb}}$ in UHV, we make use of data acquired for $t < t_\textrm{fb}$, before ring-up starts, when the particle is under steady-state feedback cooling. We convert the variance of the APD signal $\sigma(t)$ for $t<t_{\text{fb}}$ from units of \si{\volt^2} to units of \si{\kelvin} with the formula $T_{\text{fb}}=\sigma\frac{T_0}{a}\left(\frac{A'_{\text{cal}}}{A'_{\text{meas}}}\right)^2$. This value for $T_{\text{fb}}$ is then a fixed parameter in fits of Eq.~4 of the main text to ring-up data; the only free parameter is $\gamma.$

\section{Comparison of ring-up and ring-down measurements when thermal noise dominates}
\label{app:ring-up_ring-down}

\begin{figure}[ht]
	\centering
	\includegraphics[width=1\linewidth]{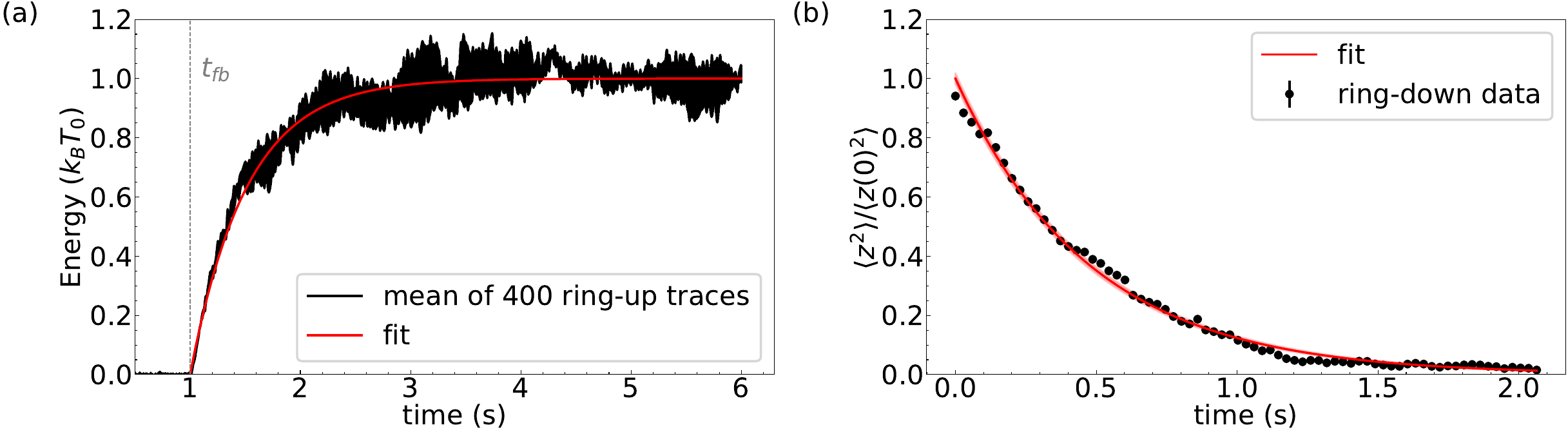}
	\caption{(a) Ring-up and (b) ring-down measurements for a second particle at pressure $P=\SI{3.1e-4}{\milli\bar}$. Error bars representing one standard deviation are plotted for the ring-down data but are smaller than the marker size. For both fits, the pink shaded region indicates 1$\sigma$ error bars on fit parameters but is only visible for the ring-down plot.}
	\label{fig:Guilio ringdown ringup}
\end{figure}
In the main text, we present a ring-up measurement at pressure $P_1 = \SI{1.2e-4}{\milli\bar}$ that provides both a damping rate at $P_1$ and a calibration between the APD voltage and the particle's energy. One expects that the same damping rate will be obtained from ring-up and ring-down measurements as long as thermal noise dominates over non-thermal noise sources, which should be the case at $P_1$. This was not verified for the specific particle described in the main text, as that particle left the Paul trap before ring-down could be measured at $P_1$. 

Here, we compare damping rates for a different particle: Fig.~\ref{fig:Guilio ringdown ringup} shows ring-up and ring-down measurement data obtained at $\SI{3.1e-4}{\milli\bar}$ with a particle with mass $m=\SI{1.5(3)e-16}{\kilo\gram}$ and charge $q=+\SI{115(22)}{\elementarycharge}$.
The data in Fig.~\ref{fig:Guilio ringdown ringup}a, which consist of the mean time trace of 400 ring-up measurements, were obtained following the same procedure as for the data in Fig.~2a in the main text and fitted using the procedure described in Appendix~\ref{app:ring-up_calibration}. As in Fig.~2a of the main text, here in Fig.~\ref{fig:Guilio ringdown ringup}a we plot the ring-up time trace already normalized to $k_B T_0$. We obtain $\gamma =2\pi\times\SI{311(8)}{\milli\hertz}$ and $T_{\text{fb}} = \SI{0.1}{\kelvin}$.

The data in Fig.~\ref{fig:Guilio ringdown ringup}b were obtained following the same procedure as for the data in Figs.~2b--d in the main text, with one minor difference: in the main text, the particle was feedback-cooled before its motion along $z$ was excited to an amplitude of hundreds of microns, while here, the particle was at room temperature before its motion was excited. 
We obtain $\gamma =2\pi\times \SI{333(5)}{\milli\hertz}$
from a fit with Eq.~1 from the main text.
As in the main text, we fit a line to the natural logarithm of the data, and the data have been normalized to the amplitude $\langle z(0)^2 \rangle$ obtained from the fit.

The two values for $\gamma$ differ by $\SI{22}{\milli\hertz}$. It is unlikely that this difference is consistent with the statistical uncertainties in the fits. We note that the pressure may have drifted slightly between the two measurements, as they took place about an hour apart. Also, the amplitude decay in Fig.~\ref{fig:Guilio ringdown ringup}b appears to have more structure than a simple exponential. (The structure cannot be attributed to stochastic kicks from thermal fluctuations, as those would be significantly smaller.) In short, the damping rates are consistent at the $\SI{20}{\percent}$ level, but further study is needed, e.g., interleaved ring-up and ring-down measurements to account for pressure drifts. 

\section{Analysis of amplitudes, uncertainties, and residuals from ring-down data}\label{appendix:residuals}
In each ring-down measurement, the particle is illuminated with a laser beam; the beam's spot size is larger than the particle's oscillation amplitude. Camera images of light scattered by the particle are taken with an exposure time much longer then the particle's oscillation period.
Each image is thus a sum of the light scattered over several particle oscillations.
For each image, we define a set of lines perpendicular to the $z$ axis, that is, perpendicular to the axis along which the particle oscillates. By adding the pixel values along each line, we generate a one-dimensional array corresponding to the intensity distribution along the particle's axis of motion.

This distribution is modeled by the convolution of the intensity distribution $P(z)$ with a Gaussian of width $w$ and maximum value $I_0$. To the convolution, a constant $c$ is added to account for  background light:
\begin{equation}
	I(z)=\int_{-\infty}^{\infty} I_0  P(z) \exp{\left[\frac{-2(z-z^\prime)^2}{w^2}\right]} dz^\prime + c
	\label{eq: model intensity profile}
\end{equation}
The piecewise-defined function $P(z)$ describes the intensity distribution of a pointlike harmonic oscillator centered at $z = z_0$ with amplitude $a$:
\begin{equation}
	P(z) = 
	\begin{cases}
		0
		& \text{if } z < (z_0-a) \text{ and } z > (z_0+a)\\
		\frac{1}{\sqrt{a^2 - (z-z_0)^2}} b(z-z_0) 
		& \text{otherwise. }
	\end{cases}
\end{equation}

\begin{figure}[ht]
		\centering
		\includegraphics[width=1\linewidth]{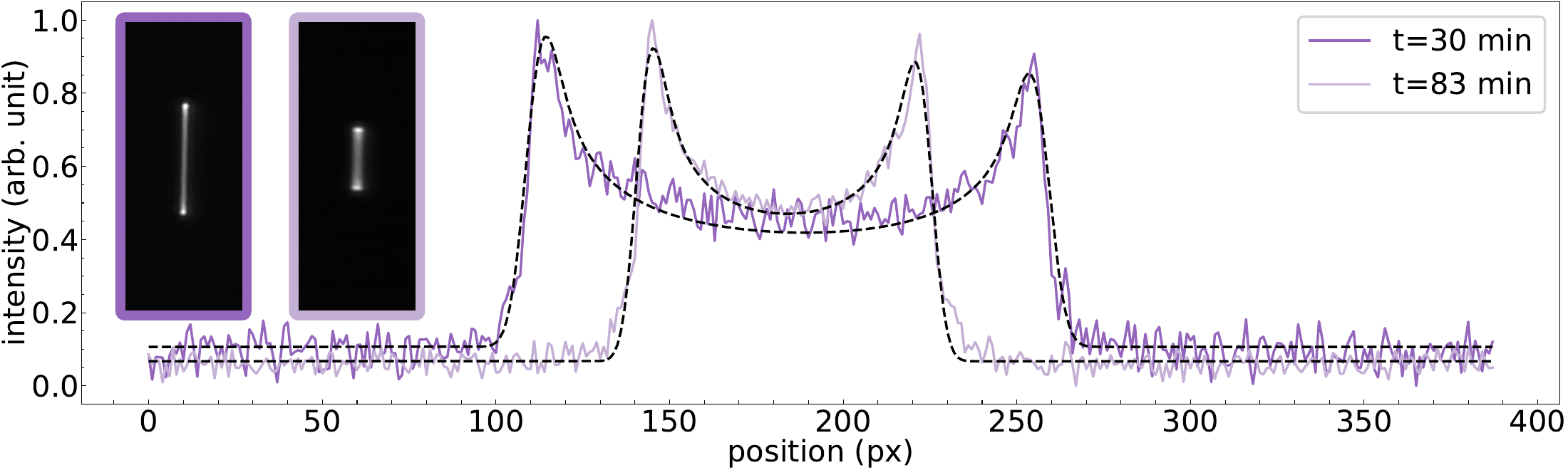}
		\caption{Examples of intensity profiles obtained from summation of camera images along lines perpendicular to the oscillation axis. The dark purple profile was recorded after 30 minutes of oscillation, the light purple profile after 83 minutes. Dashed lines are fits of Eq.~\ref{eq: model intensity profile} to the data. Parameters for the dark purple fit are $z_0=188.5(2), a=74.5(1), I_0=2.1(1), w=8.2(3), b=\SI{1.0(4)e-3}{},$ and $c=0.106(3)$. Parameters for the light purple fit are $z_0=188.027(1), a=42.0(0), I_0=1.9(1), w=6.9(3), b=\SI{0.5(3)e-3}{},$ and $c=0.067(2)$.
		Inset: corresponding camera images.}
		\label{fig:peakfinding}
\end{figure}

Here, the linear function $b(z-z_0)$ models an uneven illumination intensity, which we approximate as linear over the particle's amplitude. The Gaussian smoothing captures the fact that the particle image on the camera is not pointlike; the parameter $w$ describes the image size.
The model $I(z)$ is implemented by a numerical convolution in Python using the \texttt{numpy.convolve} function and fit to the measured intensity profiles using the \texttt{scipy.optimize.curve\_fit} function, with free parameters $I_0, z_0, w, a, b,$ and $c$.
Two examples of camera images and their corresponding intensity profiles are shown in Fig.~\ref{fig:peakfinding}, in which we see that the model reproduces the key features of the profiles. Note that the distance between the two maxima is less than $2a$ due to the Gaussian smoothing. Thus, if we were to use the distance between the maxima obtained directly from the camera data to determine the oscillation amplitude, this method would systematically underestimate $a$.

The uncertainty in the amplitude $\Delta a_\textrm{fit}$ given by the fit likely underestimates the true uncertainty: first, the other free parameters of the fit function can compensate for the amplitude uncertainty, and second, $I(z)$ is a simple model that does not account for, e.g., a detailed illumination profile of the laser.
Instead of using $\Delta a_\textrm{fit}$, we estimate the uncertainty in the amplitude $\Delta a$ based on the distance between the maxima: even though this distance underestimates $a$, a comparison between the predicted separation of the intensity profile maxima $d_{\text{model}}^{\text{max}}$ and the measured separation $d_{\text{PF}}$ nevertheless provides useful information about how well the model captures the underlying profile. 
We determine $d_\text{PF}$ from each intensity profile by taking the distance between the two major peaks that are found by the peak-finding algorithm \texttt{scipy.signal.find\_peaks}.
The uncertainty $\Delta a = \langle|d_{\text{model}}^{\text{max}} - d_{\text{PF}}| \rangle$ 
is then calculated as the mean difference between these two distances for each set of measurements.
Finally, we convert the amplitude from pixels to meters based on a calibrated camera image and find the uncertainties to be 
$\Delta a_{P_2} = \SI{3.9}{\micro\meter}$, 
$\Delta a_{P_3} = \SI{3.2}{\micro\meter}$ and
$\Delta a_{P_4} = \SI{1.9}{\micro\meter}$ for the data in Figs.~2b--d of the main text.

\begin{figure}[ht]
	\centering
	\includegraphics[width=1\linewidth]{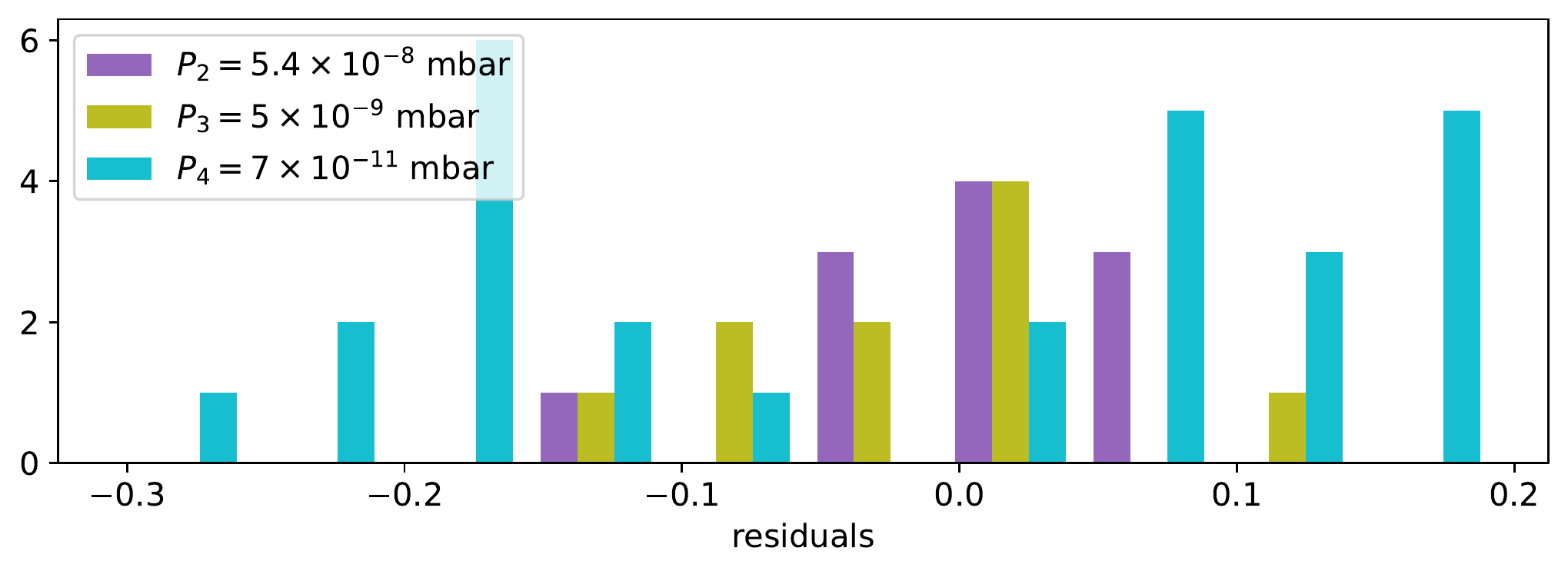}
	\caption{Histogram of the residuals of the squared amplitudes from ring-down fits at pressures $P_2$, $P_3$, and $P_4$.}
	\label{fig:residual distribution}
\end{figure}

Our analysis of ring-down data is based on the particle's energy, which is to say that the squared amplitudes, rather than the amplitudes, are the relevant quantity (see Eq.~1 of the main text). Figure~\ref{fig:residual distribution} is a histogram of the residuals $\epsilon_i$ 
of the logarithm of the squared amplitudes, calculated for all three ring-down pressures; we use the logarithm because it allows the data to be fit with a linear model.
We do not have sufficient data to draw meaningful statistical conclusions about the distribution of residuals, but we make two observations.
First, while the distribution of residuals for pressures $P_2$ and $P_3$ is centered around zero, consistent with a Gaussian distribution, the distribution for the lowest pressure $P_4$ appears bimodal. Second, as an unbiased estimator of the variance, we calculate the mean squared error (MSE), that is, the residual sum of squares divided by the number of degrees of freedom:
\begin{equation}
	\textrm{MSE} = \frac{\sum_{i = 1}^{n} \epsilon_{\textrm{norm,}i}^2}{n-p},
	\label{eq:MSE}
\end{equation}
where $n$ is the number of observations and $p = 2$ is the number of model parameters. In Eq.~\ref{eq:MSE}, we have used normalized residuals $\epsilon_{\textrm{norm,}i}$, that is, the residuals $\epsilon_i$ in the plot have been normalized by the squared amplitude of the fit. Table~1 lists mean variance and MSE for the three ring-down pressures. Here, the variance $\sigma_i^2$ for the logarithm of the squared amplitude is 
\begin{equation}
	\sigma_i^2 = \left(\frac{2\Delta a}{a_i} \right)^2,
\end{equation}
where $a_i$ is the amplitude. The mean variance is $\langle\sigma^2\rangle = \sum_{i = 1}^{n}\sigma_i^2/n$.
\begin{table}[h]
\caption{Mean variance $\langle\sigma^2\rangle$ and mean squared error (MSE) for three ring-down pressures. \label{tab:MSE}}
\begin{ruledtabular}
\begin{tabular}{ccccccc}
	Pressure & $\langle\sigma^2\rangle$   & MSE  \\ 
		\hline
	$P_2$    &0.004    &0.003 \\
	$P_3$    &0.004    &0.006 \\
	$P_4$    &0.001    &0.025 \\
\end{tabular}
\end{ruledtabular}
\end{table}
The mean variance and MSE are within a factor of 1.5 for pressures $P_2$ and $P_3$. In contrast, the MSE for $P_4$ is more than an order of magnitude larger than the mean variance, indicating that Eq.~1 does not fully capture the particle's dynamics at this pressure, as stated in the main text.

\section{Estimation of surface-induced heating}
Aside from heating due to collisions of the particle with background gas, noise on the applied voltages that generate the trapping potential, or displacement due to trap vibrations, particles levitated in an ion trap are also subject to electric-field noise due to nearby surfaces~\cite{Blatt_RFheating,martinetz2022surfaceinduced}. In this section, we estimate heating due to surfaces close to the charged particle. In the experimental setup, the closest surfaces are the radiofrequency and ground electrodes indicated in Fig.~\ref{figS1:trap_geometry}, at a distance  $d=\SI{0.9}{\milli\meter}$ from the charged particle. These stainless-steel electrodes have a diameter of $t_m = \SI{1}{\milli\meter}$ and are assumed to be at temperature $T = T_0$, at which they have a resistivity $\rho = \SI{6.9E-7}{\ohm\meter}$~\cite{lide1995crc}. For complex geometries such as those of ion traps, calculating the electric-field noise requires solving the Green's function numerically~\cite{martinetz2022surfaceinduced,kumph2016electricfield}. Nevertheless, we can estimate an upper bound on the noise due to these electrodes by treating them as infinite half-spaces, a geometry for which the electric-field noise can be calculated analytically~\cite{martinetz2022surfaceinduced,kumph2016electricfield}.  Following Eq.~14 of Ref.~\cite{kumph2016electricfield}, we find the electric-field noise at the location of the particle to be 
	\begin{equation}
	S_{EE} \approx \frac{k_\mathrm{B}T\rho}{4\pi d^3 } = \SI{3.1E-19}{\volt^2\meter^2\hertz^{-1}} \mathrm{.}    
	\end{equation}
This noise corresponds to a heating rate $\Gamma_m = 163\;\mathrm{phonons/s}$, which is more than two orders of magnitude smaller than the value of $\Gamma_\textrm{meas}^\textrm{bright}$ in the main text.
	
\section{Rotational or librational coupling as a possible noise source}

It is possible that rotational or librational modes couple to the center-of-mass motion, introducing a new dissipation channel. In this section, we consider that channel as a possible noise source. These modes will typically have much higher frequencies than the center-of-mass motion, but if we assume that the charge distribution on the nanoparticle is nonuniform (that is, if  the center of charge and center of mass are not the same), then as the particle rotates or librates, its center of charge will also rotate or librate. The particle's secular motion in the Paul trap depends on its center of charge, so the two degrees of freedom will be coupled. The dissipation channel introduced by this coupling would not affect the ring-down measurement with which we determine the quality factor because---like laser heating, electric-field noise, and displacement noise---it is a non-thermal noise source. However, it would contribute to the heating rate determined from the data in Fig.~4b of the main text. 

In the future, we plan to study this mechanism in two ways: first, by comparing reheating rates for spherical and dumbbell-shaped particles, and second, by injecting energy into a rotational mode in advance of the center-of-mass reheating measurement.
	
	\section{Measurement of vibrations in the laboratory}
	\begin{figure}[ht]
	\centering
		\includegraphics[width=0.8\linewidth]{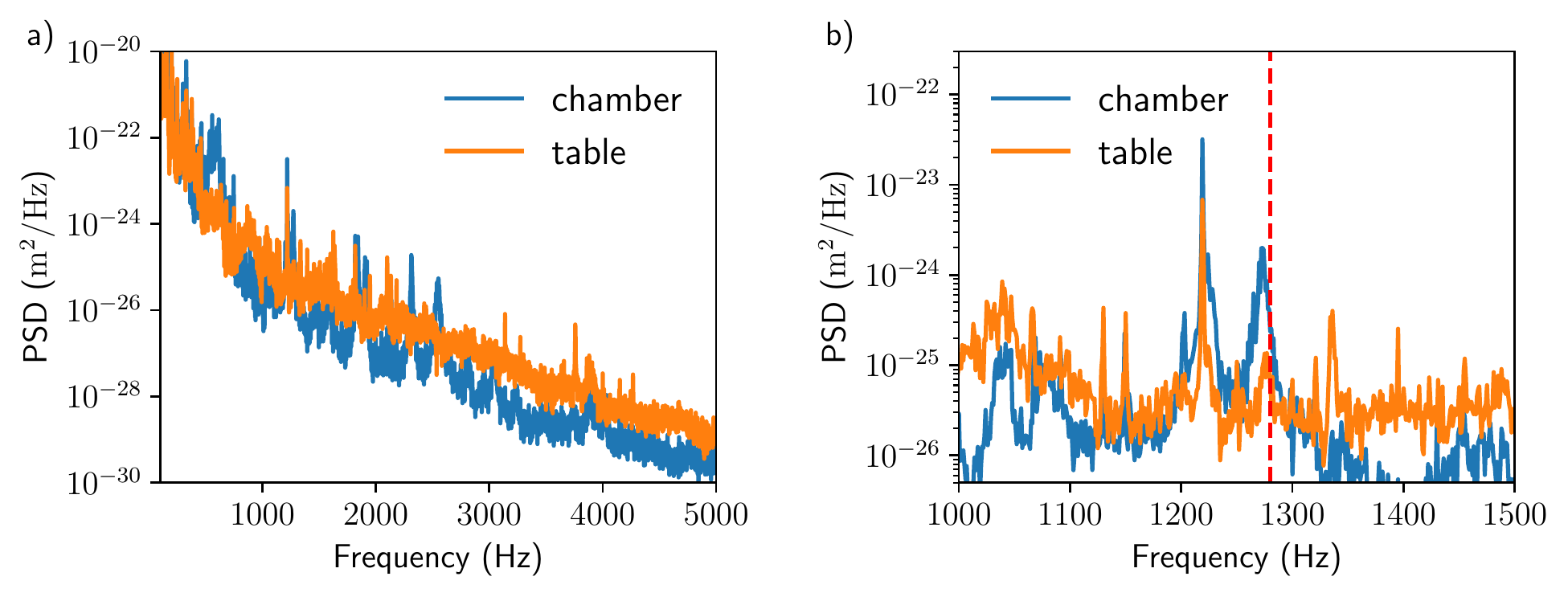} 
	\caption{ (a) Power spectral density of the displacement noise measured with an accelerometer mounted on top of the vacuum chamber (blue) and on the optical table (orange). (b) Zoomed in power spectral density of the displacement noise. Vertical red dashed line at \SI{1.28}{\kilo\hertz} indicates the resonance frequency of the nanoparticle.} 
	\label{figS6:vibration}
	\end{figure}
	We have measured the vibration level in the laboratory using a piezoelectric accelerometer (ENDEVCO 7703A-1000). Two locations were used to mount the accelerometer: on top of the vacuum chamber and on the optical table near the chamber. The accelerometer measures acceleration in the vertical direction, which is the $z$ axis in our experimental setup and coincides with the axis of particle motion in which we are interested. We convert the measured acceleration into displacement noise, the power spectral density (PSD) of which is shown in Fig.~\ref{figS6:vibration}. 
	
	First, a clear peak is visible in the vicinity of the particle's resonance frequency of \SI{1.28}{\kilo\hertz} in the PSD of the vacuum-chamber noise. In retrospect, it would have been preferable to tune the resonance frequency of the particle away from the noise peak. However, the vibrations were measured after we collected and analyzed the data presented in the main text. Second, the PSD of the noise at \SI{1.28}{\kilo\hertz} is \SI{2e-25}{\meter^2\hertz^{-1}}, which is close to the estimate obtained in the main text from the ring-up measurements at UHV (\SI{9.5e-26}{\meter^2\hertz^{-1}}). Thus, the difference between the damping rates that we infer from ring-up and ring-down measurements may be explained by displacement noise exciting the particle motion. However, we do not have a direct measurement of the acceleration at the ion trap, and it is possible that vibrations are damped between the chamber and the trap; this question thus requires further study.
	
	\section{Frequency extraction from particle time trace}
			\begin{figure}[ht]
		\centering
		\includegraphics[width=0.6\linewidth]{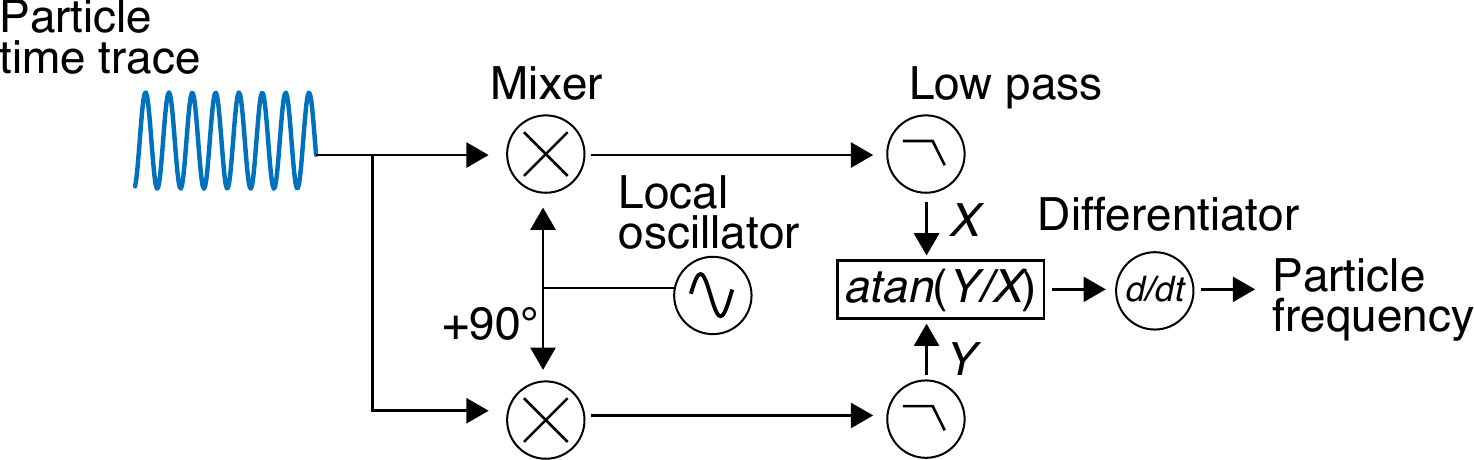}
		\caption{Schematics of the numerical code for extracting the particle's oscillation frequency from a time trace of the particle's motion.}
		\label{fig:supp_freq}
	\end{figure}
	Here, we explain how the frequency with which the particle oscillates is extracted from a time trace of the particle's motion. A phase-sensitive numerical code in MATLAB that  extracts the frequency is illustrated schematically in Fig.~\ref{fig:supp_freq}. The recorded time trace of the particle's motion along the $z$ axis is mixed with two out-of-phase local oscillators at $f_z=\SI{1.28}{\kilo\hertz}$. Each of the two mixed signals is then filtered with a low-pass filter with a cut-off frequency of \SI{5}{\hertz}. After the filters, the in-phase and out-of-phase quadratures $X$ and $Y$ are used to compute the phase quadrature of motion $\phi$ through the expression $\phi=\arctan(Y/X)$. Finally, the frequency is calculated as $f=f_z + \frac{d\phi}{dt}$.
\section{Prediction of the particle-damping coefficient from theory}
In the main text, we fit the particle damping rate $\gamma$ for four values of the pressure $P$ with the linear model $\gamma/(2\pi)=aP$, and the fit yields
a=\SI{0.9(2)}{\kilo\hertz\per\milli\bar}. 
In this section, we 
compare this value of $a$ with theory predictions. Our main limitation is that we have no direct information about the nanoparticle shape, and the shape determines the relationship between $\gamma$ and $P$. As stated in the main text, it is likely that the nanoparticle is composed of two nanospheres: the nanoparticle mass of $m=\SI{4.3e-17}{\kilo\gram}$ 
is about twice the mass of a sphere with radius $R=\SI{150}{\nano\meter}$, the nominal radius of the nanospheres we use, and also about twice the mass we typically measure for particles trapped from the same source. Here, we consider two cases: first, that the particle is spherical, and second, that the particle is a dumbbell composed of two spheres. Calculations for $\gamma$ in both cases are based on Ref.~\cite{DAHNEKE}.
\subsection{Spherical shape}
The drag force on a sphere in free molecular flow is given by Eq.~22 of Ref.~\cite{DAHNEKE}:
\begin{equation}
	F_s = - \frac{\pi}{3}\left(8 + \pi f \sqrt{T_s/T}\right)\frac{\mu r_s^2 V}{l}
	\label{eq:drag_sphere}
\end{equation}
where $f$ is the momentum accommodation factor, $T_s$ the sphere's surface temperature, $T$ the gas temperature, $\mu$ the gas viscosity, $r_s$ the sphere's radius, $V$ the relative velocity of the sphere and the gas, and $l$ the mean free path. The drag force is proportional to the velocity, $F_s = -kV$,
with spring constant $k = \gamma_s m$, so we can express the damping rate as
\begin{equation}
	\gamma_s = \frac{\pi}{3}\left(8 + \pi f \sqrt{T_s/T}\right)\frac{\mu r_s^2}{m l}
	\label{eq:damping_sphere}
\end{equation}
The viscosity can be expressed as (Eq.~20 of Ref.~\cite{DAHNEKE})
\begin{equation}
\mu = \frac{1}{2} n m_{\text{gas}} \bar{v} l,
\end{equation}
where $n$ is the number density
,  $\bar{v}=\sqrt{\frac{8k_BT}{\pi m_{\text{gas}}}}$ the mean speed of the gas molecules, and $m_{\text{gas}}$ the mass of a gas molecule, which allows us to rewrite Eq.~\ref{eq:damping_sphere}:
\begin{equation}
	\gamma_s = \frac{\pi}{6}\left(8 + \pi f \sqrt{T_s/T}\right)\frac{r_s^2 n m_{\text{gas}} \bar{v}}{m}.
\end{equation}
The ideal gas law provides us with a relationship between pressure and number density: $P = 10^{-2} n k_B T$, where $P$ is the pressure in millibar and the factor of $10^{-2}$ accounts for conversion from 
pascal to millibar. Finally, we can express $\gamma$ as a function of pressure:
\begin{align}
	\gamma_s &= \frac{10^{2}\pi}{6}\left(8 + \pi f \sqrt{T_s/T}\right)\frac{r_s^2 P m_{\text{gas}} \bar{v}}{m k_B T} \\
	&=  \frac{10^{2}}{6}\left(8 + \pi f \sqrt{T_s/T}\right)\frac{8 r_s^2 P \bar{v}}{m (8 k_B T/(\pi m_{\text{gas}}))} \\
	&=  \frac{\SI{4e2}{}}{3}\left(8 + \pi f \sqrt{T_s/T}\right)\frac{r_s^2 P}{m \bar{v}} \label{eq:gamma_sphere}
\end{align}

For the sphere's radius, we use  $r_s=2^{1/3}R$ as this value is conistent with the measured mass and the known density of silica. For the gas molecules, we assume that our UHV chamber environment is dominated by molecular hydrogen, which has a mass of $m_{\text{gas}}=\SI{3.34e-27}{\kilogram}$, and that the temperature is $T = T_0$.

The factor $f$ gives the fraction of collisions that are diffusive. For the remaining $1-f$ collisions, gas molecules undergo specular reflection from the nanoparticle. Following Ref.~\cite{DAHNEKE}, we use $f = 0.9$ as a typical value. As for the surface temperature, we assume that $T_s=T_0$ holds, that is, that the particle surface is in equilibrium with the gas. It has been demonstrated that  $T_s$ deviates from $T_0$ for nanoparticles trapped in strongly focused, high-power laser beams~\cite{millen2014nanoscale}, but the laser intensities we use for stroboscopic detection (\SI{1.5e-3}{\watt\per\centi\meter^2}) are ten orders of magnitude below those in Ref.~\cite{millen2014nanoscale}  (\SI{1.5e7}{\watt\per\centi\meter^2}). Furthermore, stroboscopic detection is intermittent: for most of the duration of each ringdown experiment, the particle is not illuminated by laser light.

With these values, Eq.~\ref{eq:gamma_sphere} predicts a damping rate $\gamma_s/(2\pi)=aP$ with $a=\SI{107(10)}{\hertz\per\milli\bar}$. Here we have only taken into account the uncertainty in our experimentally determined value for $m$ and have not assigned uncertainties to $f, T_s, T, r_s,$ or $\bar{v}$.

\subsection{Dumbbell shape}
For a dumbbell composed of two spherical particles, an analytical expression for the damping rate is not available: analytic methods are restricted to convex bodies, and Ref.~\cite{DAHNEKE} identifies ``an aggregate of several spheres" as a case for which Monte Carlo methods are required. Instead, we model the dumbbell as a cylinder with spherical ends, for which an analytical expression is available. The drag force is given by the sum of Eq.~\ref{eq:drag_sphere} and the drag force for a curved cylinder surface \cite{DAHNEKE}:
\begin{align}
	F_c &= F_s - \frac{\pi \mu L r_c V}{l}\left(f + \left(2 - \frac{6-\pi}{4}f\right)\sin^2{\theta}\right) \\
	&= - \frac{\pi}{3}\left(8 + \pi f\right)\frac{\mu r_c^2 V}{l} - \frac{\pi \mu L r_c V}{l}\left(f + \left(2 - \frac{6-\pi}{4}f\right)\sin^2{\theta}\right) \label{eq:gamma_dumbbell} 
\end{align}
where $r_\textrm{c}$ is the cylinder radius, $L$ is the cylinder length, and $\theta$ is the angle between the cylinder axis and the axis of motion relative to the gas, in this case, the $z$ axis. For Eq.~\ref{eq:gamma_dumbbell}, we have again assumed $T_{\text{sup}}=T$. 
We set the cylinder length to be $L=2r_c$, ensuring that the surface area of the cylinder is twice that of a single particle:
\begin{align}
	F_c &= - \frac{\pi}{3}\left(8 + \pi f\right)\frac{\mu r_c^2 V}{l} - \frac{2 \pi \mu r_c^2 V}{l}\left(f + \left(2 - \frac{6-\pi}{4}f\right)\sin^2{\theta}\right) \\
	&= -\pi\left(\frac{1}{3}\left(8 + \pi f\right) + 2\left(f + \left(2 - \frac{6-\pi}{4}f\right)\sin^2{\theta}\right) \right) \frac{\mu r_c^2 V}{l}
\end{align}
Following the same approach as for the sphere, we find
\begin{align}
	\gamma_c &= \frac{\pi}{2}\left(\frac{1}{3}\left(8 + \pi f\right) + 2\left(f + \left(2 - \frac{6-\pi}{4}f\right)\sin^2{\theta}\right) \right) \frac{r_c^2 n m_\textrm{gas} \bar{v}}{m} \\
	&= \SI{4e2}{}\left(\frac{1}{3}\left(8 + \pi f\right) + 2\left(f + \left(2 - \frac{6-\pi}{4}f\right)\sin^2{\theta}\right) \right) \frac{r_c^2 P}{m \bar{v}}.
	\label{eq:gamma_dumbbell2} 
\end{align}
We set the cylinder radius to be the single-particle radius $r_\textrm{c} = R$. The values of $m$, $\bar{v}$, and $f$ are the same as those in the calculation for a spherical particle. As $\theta$ is unknown and is likely to change on time scales much faster than the ringdown measurement, we average over the possible dumbbell orientations: $\sin^2\theta\to1/2$.
With these values, Eq.~\ref{eq:gamma_dumbbell2} predicts a damping rate $\gamma_c/(2\pi)=aP$ with $a=\SI{127(12)}{\hertz\per\milli\bar}$, where again only the mass uncertainty is considered.

The theory predictions for both the spherical and the dumbbell cases underestimate the value of $a$ obtained from experiment by an order of magnitude. The theory value is three standard deviations away from the experimental value: it is statistically improbable that the two values agree but cannot be ruled out. It is likely that the particle is a partially concave aggregate of two spheres that cannot be modeled analytically; to obtain a more precise comparison of theory and experiment in the future, one should trap single nanospheres (verified through measurement of the trapped particle's mass) and endeavor to reduce the uncertainty on the experimental value for $a$.

We have also made assumptions about the composition of the background gas and about temperatures of the background gas and the particle. It may be that the gas is not strictly hydrogen-dominated due to outgassing in the vicinity of the Paul-trap electrodes, and it may be that the ratio $T_s/T$ is not 1 due to heating from the driven Paul trap. However, simple estimates show that reasonable assumptions about possible gas compositions and temperatures cannot alone explain the difference between theory and experiment.

\section{Additional ring-down data at UHV}
\begin{figure}[ht]
	\centering
		\includegraphics[width=1\linewidth]{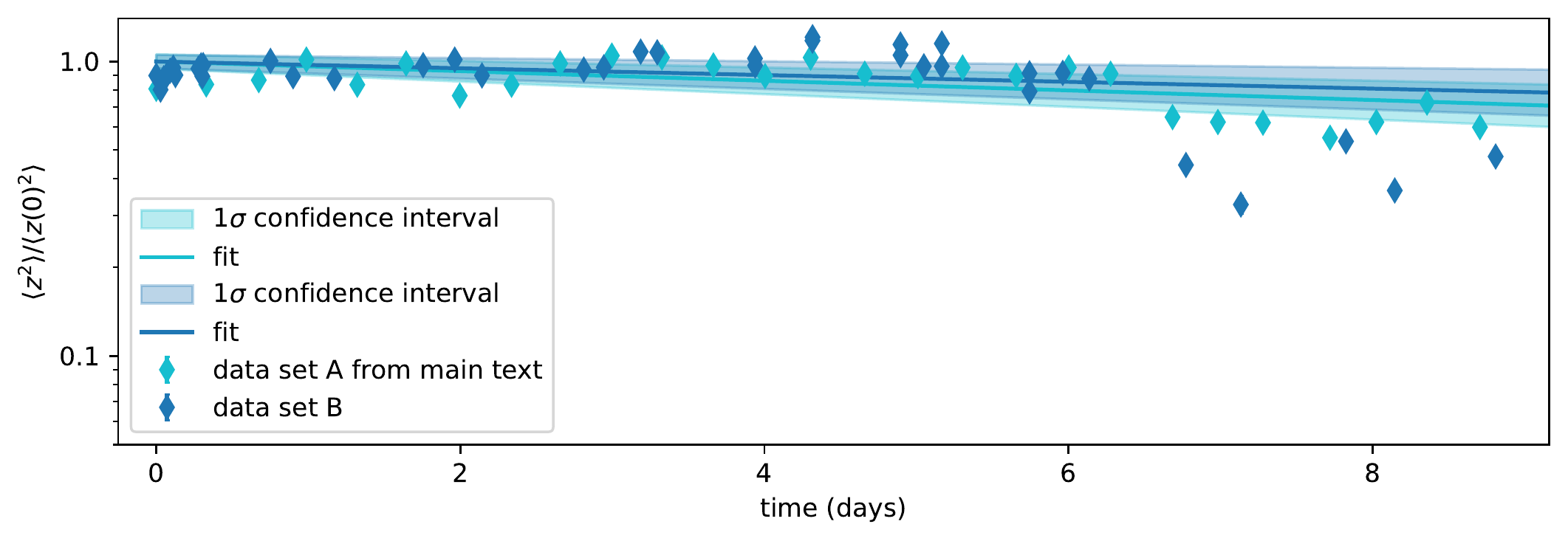} 
	\caption{Comparison between two ring-down measurements at pressure $P_4=\SI{7e-11}{\milli\bar}$. Light blue diamonds: data plotted in Fig.~2d of the main text. Dark blue diamonds: data from a second measurement, in which the particle charge and the radiofrequency drive field were both different.  
	In both cases, normalized squared amplitude is plotted as a function of time. 
	Solid lines are fits of Eq.~1 from the main text to the data. Shaded areas are 1-sigma confidence intervals. 
	}
	\label{figS7:ringdown}
\end{figure}

Here, we present results of an additional ring-down measurement (data set B) carried out at pressure $P_4=\SI{7e-11}{\milli\bar}$.
The particle is the same one that was used for all measurements presented in the main text, but it has charge $q_B=\SI{385(40)}{\elementarycharge}$, in contrast to the charge $q_A=\SI{300(30)}{\elementarycharge}$ for the measurements in the main text. 
Also, both the amplitude and frequency of the Paul-trap radiofrequency drive differ from the main text, which is reflected in the change of the $V_{\text{d}}/\Omega_{\text{d}}^2$ ratio by a factor of 0.9. Here $V_{\text{d}}$ is the amplitude and $\Omega_{\text{d}}$ is the frequency  of the Paul-trap drive. The measurement results are shown in Fig.~\ref{figS7:ringdown}. As in the main text, the data are fitted with an exponential decay function with both the decay constant $\gamma_{\text{B}}$ and the amplitude as free parameters, from which we extract  
$\gamma_{\text{B}}/(2\pi)=\SI{49(26)}{\nano\hertz}$. The fit curve and its 1-sigma confidence interval are also shown in the figure. 
The mechanical frequency of this particle is $f_z=\SI{1.45}{\kilo\hertz}$, and the quality factor is $Q= 2\pi f_z/\gamma_{\text{B}} = \SI{3(2)e10}{}$. 

For comparison, in the same plot we show the ring-down data (data set A) at $P_4$ that is presented in Fig.~2d of the main text. 
Although the data sets were obtained under different conditions, their fits yield similar damping rates. This outcome is consistent with the damping rate being determined solely by the background pressure and the particle's geometric factor $a$, both of which were held constant among the two ring-down realizations.

\bibliography{bibliography}